\documentclass[twocolumn,letterpaper,aps,prl,longbibliography,superscriptaddress,floatfix]{revtex4-2}

\usepackage{graphicx}	
\usepackage{xspace}	

\usepackage{amsmath,mathtools}
\usepackage{float}

\newcommand{\rdau}{\mbox{$R_{d{\rm Au}}$}\xspace}
\newcommand{\ypidau}{\mbox{$Y_{d{\rm Au}}^{\pi^0}$}\xspace}
\newcommand{\ydgdau}{\mbox{$Y_{d{\rm Au}}^{\gamma^{\rm dir}}$}\xspace}

\newcommand{\pt}{\mbox{$p_T$}\xspace}
\newcommand{\ptp}{\mbox{$p'_T$}\xspace}
\newcommand{\rxa}{\mbox{$R_{xA}$}\xspace}
\newcommand{\raa}{\mbox{$R_{AA}$}\xspace}
\newcommand{\rab}{\mbox{$R_{AB}$}\xspace}

\newcommand{\Ncoll}{\mbox{$N_{\rm coll}$}\xspace}

\newcommand{\pp} {\mbox{$p$$+$$p$}\xspace}
\newcommand{\dau}{\mbox{$d$$+$Au}\xspace}

\newcommand{\pau}{\mbox{$p$$+$Au}\xspace}

\newcommand{\pPb}{\mbox{$p$$+$Pb}\xspace}
\newcommand{\heau}{\mbox{$^3{\rm He}$$+$Au}\xspace}
\newcommand{\NN}{\mbox{$A$$+$$A$}\xspace}
\newcommand{\auau}{\mbox{Au$+$Au}\xspace}

\newcommand{\AB}{\mbox{$A$$+$$A$}\xspace}
\newcommand{\xA}{\mbox{$x$$+$$A$}\xspace}

\newcommand{\piz}{\mbox{$\pi^{0}$}\xspace}
\newcommand{\gdir}{\mbox{$\gamma^{\rm dir}$}\xspace}
\newcommand{\gincl}{\mbox{$\gamma^{\rm incl}_{\rm raw}$}\xspace}
\newcommand{\gev}{\mbox{GeV}\xspace}
\newcommand{\gevc}{\mbox{GeV/$c$}\xspace}

\newcommand{\Ngl}{\mbox{$N_{\rm coll}^{\rm GL}$}\xspace}
\newcommand{\Nexp}{\mbox{$N_{\rm coll}^{\rm EXP}$}\xspace}

\newcommand{\RdAuPiExp}{\mbox{$R_{d\rm Au,EXP}^{\pi^0}$}\xspace}

\newcommand{\ypipp}{\mbox{$Y_{pp}^{\pi^0}$}\xspace}
\newcommand{\ydgpp}{\mbox{$Y_{pp}^{\gamma^{\rm dir}}$}\xspace}

\newcommand{\M}{\mbox{$M(\pt,\ptp)$}\xspace}

\newcommand{\Mpigamma}{\mbox{$M(\pt(\piz),\ptp(\gamma))$}\xspace}
\newcommand{\Metagamma}{\mbox{$M(\pt(\eta),\ptp(\gamma))$}\xspace}

\begin{document}

\title{ 
Disentangling centrality bias and final-state effects in the production of 
high-$p_T$ neutral pions using direct photons in $d$$+$Au collisions at 
$\sqrt{s_{_{NN}}}=200$ GeV
}

\date{\today}

\newcommand{\abilene}{Abilene Christian University, Abilene, Texas 79699, USA}
\newcommand{\augie}{Department of Physics, Augustana University, Sioux Falls, South Dakota 57197, USA}
\newcommand{\banaras}{Department of Physics, Banaras Hindu University, Varanasi 221005, India}
\newcommand{\barc}{Bhabha Atomic Research Centre, Bombay 400 085, India}
\newcommand{\baruch}{Baruch College, City University of New York, New York, New York, 10010 USA}
\newcommand{\bnlcoll}{Collider-Accelerator Department, Brookhaven National Laboratory, Upton, New York 11973-5000, USA}
\newcommand{\bnlphys}{Physics Department, Brookhaven National Laboratory, Upton, New York 11973-5000, USA}
\newcommand{\caucr}{University of California-Riverside, Riverside, California 92521, USA}
\newcommand{\charlesczech}{Charles University, Faculty of Mathematics and Physics, 180 00 Troja, Prague, Czech Republic}
\newcommand{\cns}{Center for Nuclear Study, Graduate School of Science, University of Tokyo, 7-3-1 Hongo, Bunkyo, Tokyo 113-0033, Japan}
\newcommand{\colorado}{University of Colorado, Boulder, Colorado 80309, USA}
\newcommand{\columbia}{Columbia University, New York, New York 10027 and Nevis Laboratories, Irvington, New York 10533, USA}
\newcommand{\czechtech}{Czech Technical University, Zikova 4, 166 36 Prague 6, Czech Republic}
\newcommand{\debrecen}{Debrecen University, H-4010 Debrecen, Egyetem t{\'e}r 1, Hungary}
\newcommand{\elte}{ELTE, E{\"o}tv{\"o}s Lor{\'a}nd University, H-1117 Budapest, P{\'a}zm{\'a}ny P.~s.~1/A, Hungary}
\newcommand{\ewha}{Ewha Womans University, Seoul 120-750, Korea}
\newcommand{\famu}{Florida A\&M University, Tallahassee, FL 32307, USA}
\newcommand{\fsu}{Florida State University, Tallahassee, Florida 32306, USA}
\newcommand{\gsu}{Georgia State University, Atlanta, Georgia 30303, USA}
\newcommand{\hiroshima}{Physics Program and International Institute for Sustainability with Knotted Chiral Meta Matter (WPI-SKCM$^2$), Hiroshima University, Higashi-Hiroshima, Hiroshima 739-8526, Japan}
\newcommand{\howard}{Department of Physics and Astronomy, Howard University, Washington, DC 20059, USA}
\newcommand{\hunrenatomki}{HUN-REN ATOMKI, H-4026 Debrecen, Bem t{\'e}r 18/c, Hungary}
\newcommand{\ihepprot}{IHEP Protvino, State Research Center of Russian Federation, Institute for High Energy Physics, Protvino, 142281, Russia}
\newcommand{\illuiuc}{University of Illinois at Urbana-Champaign, Urbana, Illinois 61801, USA}
\newcommand{\inrras}{Institute for Nuclear Research of the Russian Academy of Sciences, prospekt 60-letiya Oktyabrya 7a, Moscow 117312, Russia}
\newcommand{\instpasczech}{Institute of Physics, Academy of Sciences of the Czech Republic, Na Slovance 2, 182 21 Prague 8, Czech Republic}
\newcommand{\isu}{Iowa State University, Ames, Iowa 50011, USA}
\newcommand{\jaea}{Advanced Science Research Center, Japan Atomic Energy Agency, 2-4 Shirakata Shirane, Tokai-mura, Naka-gun, Ibaraki-ken 319-1195, Japan}
\newcommand{\jeonbuk}{Jeonbuk National University, Jeonju, 54896, Korea}
\newcommand{\jyvaskyla}{Helsinki Institute of Physics and University of Jyv{\"a}skyl{\"a}, P.O.Box 35, FI-40014 Jyv{\"a}skyl{\"a}, Finland}
\newcommand{\kek}{KEK, High Energy Accelerator Research Organization, Tsukuba, Ibaraki 305-0801, Japan}
\newcommand{\korea}{Korea University, Seoul 02841, Korea}
\newcommand{\kurchatov}{National Research Center ``Kurchatov Institute", Moscow, 123098 Russia}
\newcommand{\kyoto}{Kyoto University, Kyoto 606-8502, Japan}
\newcommand{\losalamos}{Los Alamos National Laboratory, Los Alamos, New Mexico 87545, USA}
\newcommand{\lund}{Department of Physics, Lund University, Box 118, SE-221 00 Lund, Sweden}
\newcommand{\lyon}{IPNL, CNRS/IN2P3, Univ Lyon, Universit{\'e} Lyon 1, F-69622, Villeurbanne, France}
\newcommand{\maryland}{University of Maryland, College Park, Maryland 20742, USA}
\newcommand{\mate}{MATE, Institute of Technology, Laboratory of Femtoscopy, K\'aroly R\'obert Campus, H-3200 Gy\"ongy\"os, M\'atrai \'ut 36, Hungary}
\newcommand{\michigan}{Department of Physics, University of Michigan, Ann Arbor, Michigan 48109-1040, USA}
\newcommand{\miss}{Mississippi State University, Mississippi State, Mississippi 39762, USA}
\newcommand{\muhlenberg}{Muhlenberg College, Allentown, Pennsylvania 18104-5586, USA}
\newcommand{\nara}{Nara Women's University, Kita-uoya Nishi-machi Nara 630-8506, Japan}
\newcommand{\natmephi}{National Research Nuclear University, MEPhI, Moscow Engineering Physics Institute, Moscow, 115409, Russia}
\newcommand{\newmex}{University of New Mexico, Albuquerque, New Mexico 87131, USA}
\newcommand{\nmsu}{New Mexico State University, Las Cruces, New Mexico 88003, USA}
\newcommand{\northcg}{Physics and Astronomy Department, University of North Carolina at Greensboro, Greensboro, North Carolina 27412, USA}
\newcommand{\ohio}{Department of Physics and Astronomy, Ohio University, Athens, Ohio 45701, USA}
\newcommand{\ornl}{Oak Ridge National Laboratory, Oak Ridge, Tennessee 37831, USA}
\newcommand{\orsay}{IPN-Orsay, Univ.~Paris-Sud, CNRS/IN2P3, Universit\'e Paris-Saclay, BP1, F-91406, Orsay, France}
\newcommand{\peking}{Peking University, Beijing 100871, People's Republic of China}
\newcommand{\pnpi}{PNPI, Petersburg Nuclear Physics Institute, Gatchina, Leningrad region, 188300, Russia}
\newcommand{\pusan}{Pusan National University, Pusan 46241, Korea}
\newcommand{\riken}{RIKEN Nishina Center for Accelerator-Based Science, Wako, Saitama 351-0198, Japan}
\newcommand{\rikjrbrc}{RIKEN BNL Research Center, Brookhaven National Laboratory, Upton, New York 11973-5000, USA}
\newcommand{\rikkyo}{Physics Department, Rikkyo University, 3-34-1 Nishi-Ikebukuro, Toshima, Tokyo 171-8501, Japan}
\newcommand{\saispbstu}{Saint Petersburg State Polytechnic University, St.~Petersburg, 195251 Russia}
\newcommand{\seoulnat}{Department of Physics and Astronomy, Seoul National University, Seoul 151-742, Korea}
\newcommand{\stonybrkc}{Chemistry Department, Stony Brook University, SUNY, Stony Brook, New York 11794-3400, USA}
\newcommand{\stonycrkp}{Department of Physics and Astronomy, Stony Brook University, SUNY, Stony Brook, New York 11794-3800, USA}
\newcommand{\tenn}{University of Tennessee, Knoxville, Tennessee 37996, USA}
\newcommand{\texsu}{Texas Southern University, Houston, TX 77004, USA}
\newcommand{\titech}{Department of Physics, Tokyo Institute of Technology, Oh-okayama, Meguro, Tokyo 152-8551, Japan}
\newcommand{\tsukuba}{Tomonaga Center for the History of the Universe, University of Tsukuba, Tsukuba, Ibaraki 305, Japan}
\newcommand{\usmma}{United States Merchant Marine Academy, Kings Point, New York 11024, USA}
\newcommand{\vandy}{Vanderbilt University, Nashville, Tennessee 37235, USA}
\newcommand{\weizmann}{Weizmann Institute, Rehovot 76100, Israel}
\newcommand{\wigner}{Institute for Particle and Nuclear Physics, HUN-REN Wigner Research Centre for Physics, (HUN-REN Wigner RCP, RMI), H-1525 Budapest 114, POBox 49, Budapest, Hungary}
\newcommand{\yonsei}{Yonsei University, IPAP, Seoul 120-749, Korea}
\newcommand{\zagreb}{Department of Physics, Faculty of Science, University of Zagreb, Bijeni\v{c}ka c.~32 HR-10002 Zagreb, Croatia}
\newcommand{\zambia}{Department of Physics, School of Natural Sciences, University of Zambia, Great East Road Campus, Box 32379, Lusaka, Zambia}
\affiliation{\abilene}
\affiliation{\augie}
\affiliation{\banaras}
\affiliation{\barc}
\affiliation{\baruch}
\affiliation{\bnlcoll}
\affiliation{\bnlphys}
\affiliation{\caucr}
\affiliation{\charlesczech}
\affiliation{\cns}
\affiliation{\colorado}
\affiliation{\columbia}
\affiliation{\czechtech}
\affiliation{\debrecen}
\affiliation{\elte}
\affiliation{\ewha}
\affiliation{\famu}
\affiliation{\fsu}
\affiliation{\gsu}
\affiliation{\hiroshima}
\affiliation{\howard}
\affiliation{\hunrenatomki}
\affiliation{\ihepprot}
\affiliation{\illuiuc}
\affiliation{\inrras}
\affiliation{\instpasczech}
\affiliation{\isu}
\affiliation{\jaea}
\affiliation{\jeonbuk}
\affiliation{\jyvaskyla}
\affiliation{\kek}
\affiliation{\korea}
\affiliation{\kurchatov}
\affiliation{\kyoto}
\affiliation{\losalamos}
\affiliation{\lund}
\affiliation{\lyon}
\affiliation{\maryland}
\affiliation{\mate}
\affiliation{\michigan}
\affiliation{\miss}
\affiliation{\muhlenberg}
\affiliation{\nara}
\affiliation{\natmephi}
\affiliation{\newmex}
\affiliation{\nmsu}
\affiliation{\northcg}
\affiliation{\ohio}
\affiliation{\ornl}
\affiliation{\orsay}
\affiliation{\peking}
\affiliation{\pnpi}
\affiliation{\pusan}
\affiliation{\riken}
\affiliation{\rikjrbrc}
\affiliation{\rikkyo}
\affiliation{\saispbstu}
\affiliation{\seoulnat}
\affiliation{\stonybrkc}
\affiliation{\stonycrkp}
\affiliation{\tenn}
\affiliation{\texsu}
\affiliation{\titech}
\affiliation{\tsukuba}
\affiliation{\usmma}
\affiliation{\vandy}
\affiliation{\weizmann}
\affiliation{\wigner}
\affiliation{\yonsei}
\affiliation{\zagreb}
\affiliation{\zambia}
\author{N.J.~Abdulameer} \affiliation{\debrecen} \affiliation{\hunrenatomki}
\author{U.~Acharya} \affiliation{\gsu} 
\author{C.~Aidala} \affiliation{\michigan} 
\author{Y.~Akiba} \email[PHENIX Spokesperson: ]{akiba@rcf.rhic.bnl.gov} \affiliation{\riken} \affiliation{\rikjrbrc} 
\author{M.~Alfred} \affiliation{\howard} 
\author{K.~Aoki} \affiliation{\kek} 
\author{N.~Apadula} \affiliation{\isu} 
\author{C.~Ayuso} \affiliation{\michigan} 
\author{V.~Babintsev} \affiliation{\ihepprot} 
\author{K.N.~Barish} \affiliation{\caucr} 
% ...
\collaboration{PHENIX Collaboration}  \noaffiliation


\vspace{1.0cm}

\begin{abstract}


PHENIX presents a simultaneous measurement of the production of direct 
$\gamma$ and $\pi^{0}$ in $d$$+$Au collisions at $\sqrt{s_{_{NN}}}=200$ 
GeV over a $p_T$ range of 7.5 to 18 GeV/$c$ for different event samples 
selected by event activity, i.e. charged-particle multiplicity detected 
at forward rapidity. Direct-photon yields are used to empirically 
estimate the contribution of hard-scattering processes in the different 
event samples. Using this estimate, the average nuclear-modification 
factor, $R_{d\rm Au,EXP}^{\pi^0}$, is $0.925{\pm}0.023{\rm 
(stat)}{\pm}0.15{\rm (scale)}$, consistent with unity for minimum-bias 
(MB) $d$$+$Au collisions. For event classes with low and moderate event 
activity, $R_{d\rm Au,EXP}^{\pi^0}$ is consistent with the MB value 
within 5\% uncertainty. This result confirms that the previously 
observed enhancement of high-$p_T$~$\pi^0$ production found in 
small-system collisions with low event activity is a result of a bias in 
interpreting event activity within the Glauber framework. In contrast, 
for the top 5\% of events with the highest event activity, 
$R_{d\rm Au,EXP}^{\pi^0}$ is suppressed by 20\% relative to the MB value 
with a significance of $4.5\sigma$, which may be due to final-state 
effects. This suppression corresponds to a $p_T$ shift of 
$\delta{p_T}=0.213\pm0.055$~Gev/$c$ at 9~Gev/$c$.

\end{abstract}

\maketitle



High transverse-momentum (\pt) particles are produced in rare, initial 
hard-scattering processes and are sensitive to the evolution of relativistic 
heavy-ion collisions. The suppression of their yields with respect to the 
incoherent superposition of yields from \pp collisions was 
predicted~\cite{Wang:1992qdg,Wang:1998bha,Gyulassy:2000fs} as a signature 
for the formation of a hot and dense partonic medium, the quark-gluon plasma 
(QGP). This was first observed in \AB collisions at the Relativistic Heavy Ion 
Collider (RHIC)~\cite{PHENIX:2001hpc,PHENIX:2008saf} and later at the Large 
Hadron Collider (LHC)~\cite{ALICE:2012aqc,ATLAS:2015qmb,CMS:2015ved}. 
Together with the absence of suppression in minimum-bias (MB) 
\dau~\cite{PHENIX:2003qdw,STAR:2003pjh} collisions at RHIC and \pPb 
collisions at the LHC~\cite{ALICE:2012mj,CMS:2016xef}, it served as a 
compelling piece of evidence that QGP is formed in heavy-ion collisions.

Multiparticle 
correlations~\cite{ALICE:2014dwt,PHENIX:2014fnc,CMS:2015yux,CMS:2016fnw,ALICE:2017svf,PHENIX:2018lia} 
and strangeness enhancement~\cite{ALICE:2016fzo} in collisions of 
small-on-large nuclei (\xA) with high particle multiplicity, or event 
activity, have led to the suggestion that QGP droplets may be formed even in 
small systems. If true, one may also find evidence for energy loss of 
high-\pt particles in these collisions. However, measurements at 
RHIC~\cite{Wysocki:2013caa,PHENIX:2021dod} and 
LHC~\cite{ATLAS:2014cpa,ATLAS:2016xpn} have revealed an inconclusive pattern 
of suppression in high-activity events, and a puzzling enhancement in 
low-activity events.

Theoretical calculations predict that any presence of QGP in \xA collisions 
should result in a suppression of high-\pt 
hadrons~\cite{Huss:2020whe,Ke:2022gkq}. While there are now stringent 
experimental limits on energy loss for jets with \pt above 15 \gevc in 
$p+$Pb collisions at the LHC~\cite{ALICE:2017svf,ATLAS:2022iyq}, the \pt 
range below 15 \gevc remains less constrained, and the cause of the 
enhancement in low activity events remains unclear. To better understand 
these observations, high \pt particle production in \xA at RHIC needs to be 
explored with greater accuracy.

Evidence for energy loss is typically quantified by the nuclear-modification 
factor, \rab, as a function of \pt:
\begin{equation}\label{eq:RAB}
R_{AB}(\pt) = \frac{Y_{AB}(\pt) }
{N_{\rm coll} \, Y_{pp}(\pt)},
\end{equation}
\noindent where $Y_{AB}$ and $Y_{pp}$ are the yields in $A$$+$$B$ and \pp 
collisions, respectively, with $A$ and $B$ being large or small ions. The 
average number of binary nucleon-nucleon ($NN$) collisions, \Ncoll is used 
to scale particle production from hard-scattering processes from \pp to 
$A$$+$$B$ events. Because \Ncoll is not experimentally accessible, the Glauber 
model (GLM)~\cite{Glauber:1970jm,Miller:2007ri} is usually used to map 
\Ncoll to the measured event activity or centrality. The basic tenet is that 
the majority of $NN$ collisions involve only small momentum exchanges; thus, 
\Ncoll can be estimated with the eikonal approximation.

The observation that the direct photon \raa is consistent with unity in 
\auau collisions, independent of the event selection~\cite{PHENIX:2012jbv}, 
confirmed that the particle production from hard-scattering processes scales 
with \Ncoll. Similar behavior has been seen at the LHC for electromagnetic 
(EM) 
probes~\cite{CMS:2012oiv,ALICE:2015xmh,ATLAS:2015rlt,ATLAS:2019maq,CMS:2021kvd} 
including the Z boson~\footnote{There is some disagreement between ATLAS and 
CMS in event classes with low activity where \Ncoll is small and the events 
are categorized in a different way.}.

Studies at RHIC and 
LHC~\cite{PHENIX:2013jxf,ALICE:2014xsp,Kordell:2016njg} indicate that 
the GLM based mapping of various measures of event activity to \Ncoll 
can be biased by the presence of hard-scattering processes. The effect 
will not be noticeable if \Ncoll is large~\cite{ALICE:2014xsp}, but it 
can be significant if \Ncoll is small, as in peripheral \NN collisions 
or collisions of \xA 
systems~\cite{David:2014zya,Kordell:2016njg,Loizides:2017sqq,ALICE:2018ekf,Bzdak:2014rca}. 
In \xA collisions, this bias would manifest itself as an underestimate 
(overestimate) of \Ncoll for events with low (high) event activity 
leading to an apparent enhancement (suppression) of high-\pt hadron or 
jet yields. Although the effect of this selection bias has been studied 
extensively~\cite{Alvioli:2013vk,Alvioli:2014eda,McGlinchey:2016ssj}, it 
remains a challenge to disentangle the final-state effects in \rxa from 
the impact of this bias.

This letter aims to resolve the ambiguity of whether the observed 
enhancement and/or suppression pattern in \rdau for \dau collisions 
selected by event activity~\cite{PHENIX:2021dod} is due to an 
event-selection bias in estimating \Ncoll or true nuclear effects. To 
achieve this, high-\pt direct photons (\gdir) are employed as a benchmark 
for particle production from hard-scattering processes in a given event 
sample~\cite{Reygers:2004}. They are used to experimentally estimate the 
number of binary collisions (\Nexp) for a given event selection from the 
ratio of the direct-photon yields in that selection to that from \pp 
collisions:

\begin{equation}\label{eq:Ncoll_exp}
 \Nexp(\pt) = \frac{\ydgdau(\pt)}{\ydgpp(\pt)}. 
\end{equation}

Here it is assumed that final-state effects on photons are negligible.  
Indeed, in \auau collisions, where $R_{AA}^{\gamma^{{\rm dir}}}$ is 
consistent with unity and shows no appreciable \pt 
dependence~\cite{PHENIX:2012jbv}, \Nexp is equal to \Ncoll as determined by 
the GLM (\Ngl).  Because cold-nuclear-matter (CNM) effects on \gdir in \dau 
are expected to be similar or smaller than in \auau~\cite{Arleo:2011gc}, 
\Nexp is also a measure of \Ncoll in \dau with the advantage that it is less 
sensitive to potential event-selection biases than \Ngl. Theoretical 
calculations suggest that there are changes in the probability of hard 
scattering owing to the presence of CNM effects, including differences in 
isospin, i.e. different u and d quark content in \pp and \dau collisions, 
shadowing, the EMC effect, etc.~\cite{Arleo:2011gc,Zhang:2008ek,Ke:2022gkq}. 
These are predicted to result in a reduction in the production of high \pt 
\gdir in \dau collisions of up to 10\% over the \pt range investigated 
here~\cite{Arleo:2011gc}. The same calculations show a similar decrease in 
pion production. Accounting for the different Bjorken-$x$ regions spanned by 
\gdir and \piz, the \pt dependence of their relative yields between \dau and 
\pp collisions cancels within 5\%. While these calculations are for MB 
collisions, they are expected to hold true for all event selections. 
Compared to the current experimental uncertainties these differences are 
small, further justifying the use of Eq.~\ref{eq:Ncoll_exp} to test the 
scaling of high-\pt particle production from \pp to a given \dau event 
sample.


The \piz and direct-photon data were recorded with the PHENIX 
experiment~\cite{PHENIX:2003nhg} in 2016 using a triggered event 
sample of 12.6{$\times$}10$^6$ \dau collisions, corresponding to an 
integrated luminosity of ${\approx}50$ nb$^{-1}$. In addition, a MB data 
sample of 65{$\times$}10$^6$ events is used to define event activity 
classes, based on the charged particle multiplicity in the Au-going 
direction, and to determine the absolute normalization of the \piz and 
direct photon spectra. The MB trigger requires a coincidence of at least one 
hit in the upstream and downstream beam-beam Counters~\cite{PHENIX:2003tlh} 
(covering the pseudorapidity range $3.0<|\eta|<3.9$), which records 
$88\pm4$\% of the inelastic cross section. The event trigger required a 
local energy deposit ($>$2.4\,\gev) in the electromagnetic 
calorimeter~\cite{PHENIX:2003fvo} (EMCal, $|\eta|<0.35$).

The \piz mesons are reconstructed using the 
$\pi^0\rightarrow\gamma\gamma$ decay as described 
in~\cite{PHENIX:2008saf,PHENIX:2012jha}. Photon candidates are identified 
by comparing the shape and timing of the reconstructed energy clusters to 
the expected response of the EMCal. All photon candidates in an event are 
combined into pairs, their invariant mass is calculated, and the mass 
distribution is aggregated in bins of reconstructed \ptp. Any 
combinatorial background is subtracted. The result is the raw \piz yield, 
$dN^{\pi^0}_{\rm raw}/d\ptp$, which is corrected to represent the true 
\piz yield, $dN^{\pi^0}/d\pt$, in the rapidity range $|y|<0.5$.  The 
correction is determined by an iterative unfolding procedure similar to 
the standard Bayesian approach, using a response matrix, \M, with 
elements that are the probability that a \piz of a given true \pt, 
uniformly distributed in azimuth and in $|y|<0.5$, will be reconstructed 
in PHENIX as a \piz with \ptp. The simulated response is determined with 
a GEANT3~\cite{Brun:1994aa} implementation of the PHENIX experiment. The 
yield of photon candidates constitutes the raw inclusive photon yield, 
\gincl, and contains energy clusters from direct photons, clusters from 
single decay photons, and clusters reconstructed from overlapping EM 
showers from two decay photons. The direct-photon yield is extracted from 
the \gincl yield following~\cite{PHENIX:2008saf,PHENIX:2012jbv} without 
isolation requirements. To determine the contribution of decay photons to 
the raw inclusive yield, two additional response matrices are generated, 
\Mpigamma and \Metagamma. Here the elements are the number of photon 
candidates reconstructed with \ptp for a \piz ($\eta$) of a given \pt. 
The reconstructed decay photon candidates from \piz are given by 
$M(\pt(\piz),\ptp(\gamma)) \times dN^{\pi^0}/d\pt$, using the corrected 
\piz \pt spectrum.

The contribution from other meson decays is calculated as 
$M(\pt(\eta),\ptp(\gamma)) \times dN^{\eta}/d\pt \times 
(1+\gamma^{\omega,\eta'}/\gamma^\eta)$. The $\eta$ meson \pt spectrum is 
taken as $dN^{\eta}/d\pt= \eta/\piz \times dN^{\pi^0}/d\pt$, using the 
\pt dependent $\eta/\piz$ ratio determined in~\cite{Ren:2021xbh}. The 
photon candidates from $\eta$ decays are then scaled by the ratio 
$\gamma^{\omega,\eta'}/\gamma^\eta = 0.19$ to account for the 
contribution from $\omega$ and $\eta'$ meson decays, which is independent 
of \pt above 7~\gevc~\cite{PHENIX:2014nkk}. This raw decay-photon 
contribution to the photon candidates is subtracted from the raw 
inclusive-photon yield. The remaining photon candidates constitute the 
raw direct-photon yield, which is corrected using the same iterative 
method deployed for \piz~\cite{Ram:2021}. Finally, for each event class, 
the absolute normalization of \piz and \gdir yields is corrected for the 
average \pt independent bias induced by a hard process on 
\Ngl~\cite{PHENIX:2013jxf}.

The systematic uncertainties on the \piz and \gdir yields are evaluated 
following established procedures~\cite{PHENIX:2021dod}. For \dau they vary 
between 11\% to 15\% over the \pt range from 7.5 to 18 \gevc. Sources of 
uncertainties include the energy scale calibration, the amount of material 
in the detector where photons convert, the photon shower identification, 
shower merging, the absolute normalization, plus other smaller 
contributions. These sources are common to \piz and \gdir. For \gdir 
additional uncertainties due to the hadron contamination and the 
contribution of decay photons from $\eta$, $\omega$, and $\eta'$ are 
considered. In the \gdir/\piz ratio many uncertainties cancel, including 
uncertainties from normalization, reducing the systematic uncertainties to 
about 6\% to 9\% independent of event activity. To assure that there are no 
event-activity-dependent uncertainties, the corrections for the \piz and 
\gdir yields were determined separately for each event class. With exception 
of the absolute normalization correction, they are found to be equal within 
an accuracy of 1\% 
(see supplemental material~\footnote{See Supplemental Material at [URL 
will be inserted by publisher], which includes 
Refs.~\cite{Aguilar:2021sfa,PHENIX:2015vqa,PISA}, for detailed 
explanations for the specialists on breakdown of the systematic 
uncertainties, their correlations and propagation, as well as cancellation 
in ratios.}).

\begin{figure}[!htbp] 
        \includegraphics[width=1.0\linewidth]{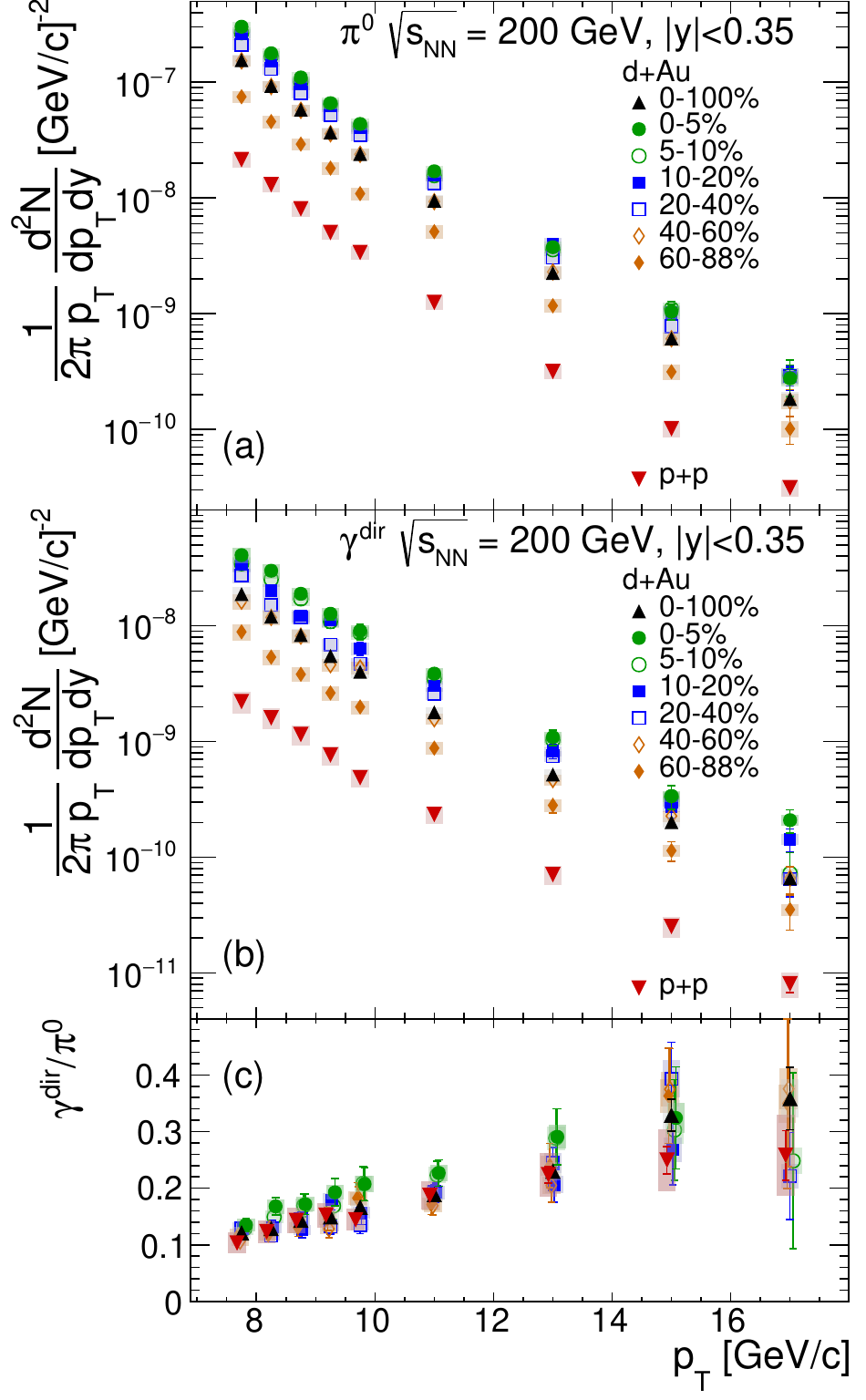}
\caption{The \pt distribution at high \pt of (a) neutral pions and 
(b) direct photons for different \dau event activity classes compared to 
those from \pp collisions. Panel (c) shows the ratio $\gdir/\piz$.  For 
better visibility the points are slightly shifted in \pt.
        }
\label{fig:spectra} 
\end{figure}

Invariant yields of \piz and \gdir covering the \pt range from 7.5 to 18 
\gevc are shown in Fig.~\ref{fig:spectra}(a) and (b), respectively. 
Both panels include the yield for \dau (0\%-100\%) and for six \dau 
event classes selected by event activity, with 0\%--5\% being the events 
with the largest activity. Invariant yields measured in 
\pp~\cite{PHENIX:2012jgx,PHENIX:2021dod} are also shown. The \dau 
results for \piz and for MB \gdir are consistent with previous 
measurements~\cite{PHENIX:2021dod, PHENIX:2012krx}. 
Fig.~\ref{fig:spectra}(c) presents the \gdir/\piz ratios. The 
\gdir/\piz ratio for \dau (0\%--100\%) is consistent with that from 
\pp collisions. This is also true for all \dau event classes with low to 
moderate event activity. The similarity of \gdir/\piz for \pp and most 
\dau collisions suggests that initial state CNM effects must be similar
for the production of high-\pt \piz and \gdir. This supports the 
conjecture that the earlier observed enhancement of $R_{xA}^{\piz}$ in 
\xA collisions with low event activity~\cite{PHENIX:2021dod} was  caused 
by a bias in the mapping of event activity to \Ngl. In contrast, the 
\gdir/\piz ratio for the \dau events with high activity (0\%--5\%) is 
visibly larger than the one for \pp.

\begin{figure*}[!htbp] 
\includegraphics[width=0.99\linewidth]{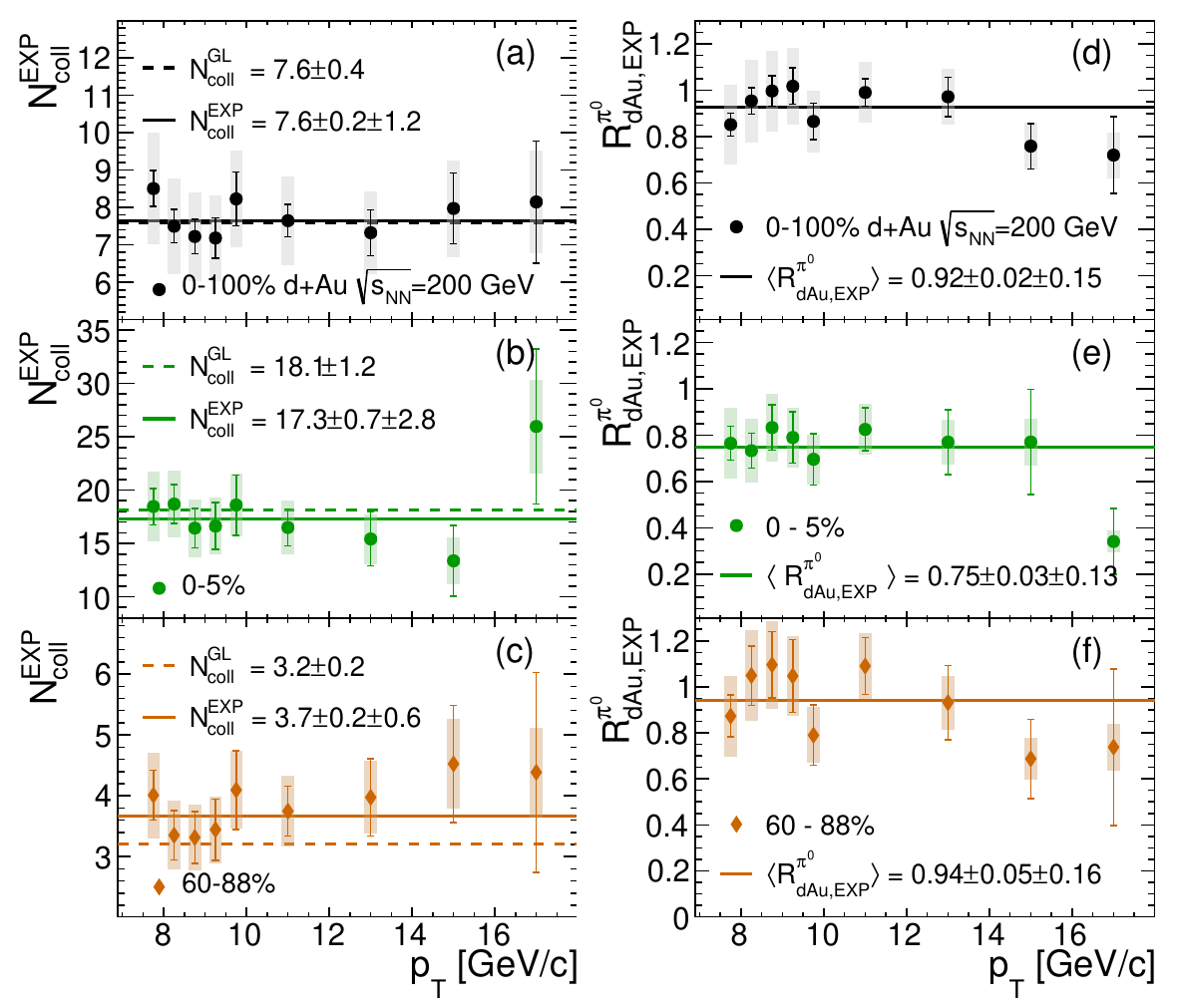}
\caption{Values of \Nexp versus \pt as defined in Eq.~\ref{eq:Ncoll_exp} for 
three \dau event classes, (a) 0\%--100\%, (b) 0\%--5\% and (c) 60\%--88\%. 
Also shown are fits to the data (solid lines) and the corresponding value 
\Ngl (dashed lines).  Panels (d) to (f) show the nuclear modification 
factors \RdAuPiExp, calculated with Eq.~\ref{eq:double_ratio}, for the same 
event selections as in panels (a) to (c) together with fits to the data. }
\label{fig:rdau_pt}
\end{figure*}

\begin{figure}[!htb] 
        \includegraphics[width=1.0\linewidth]{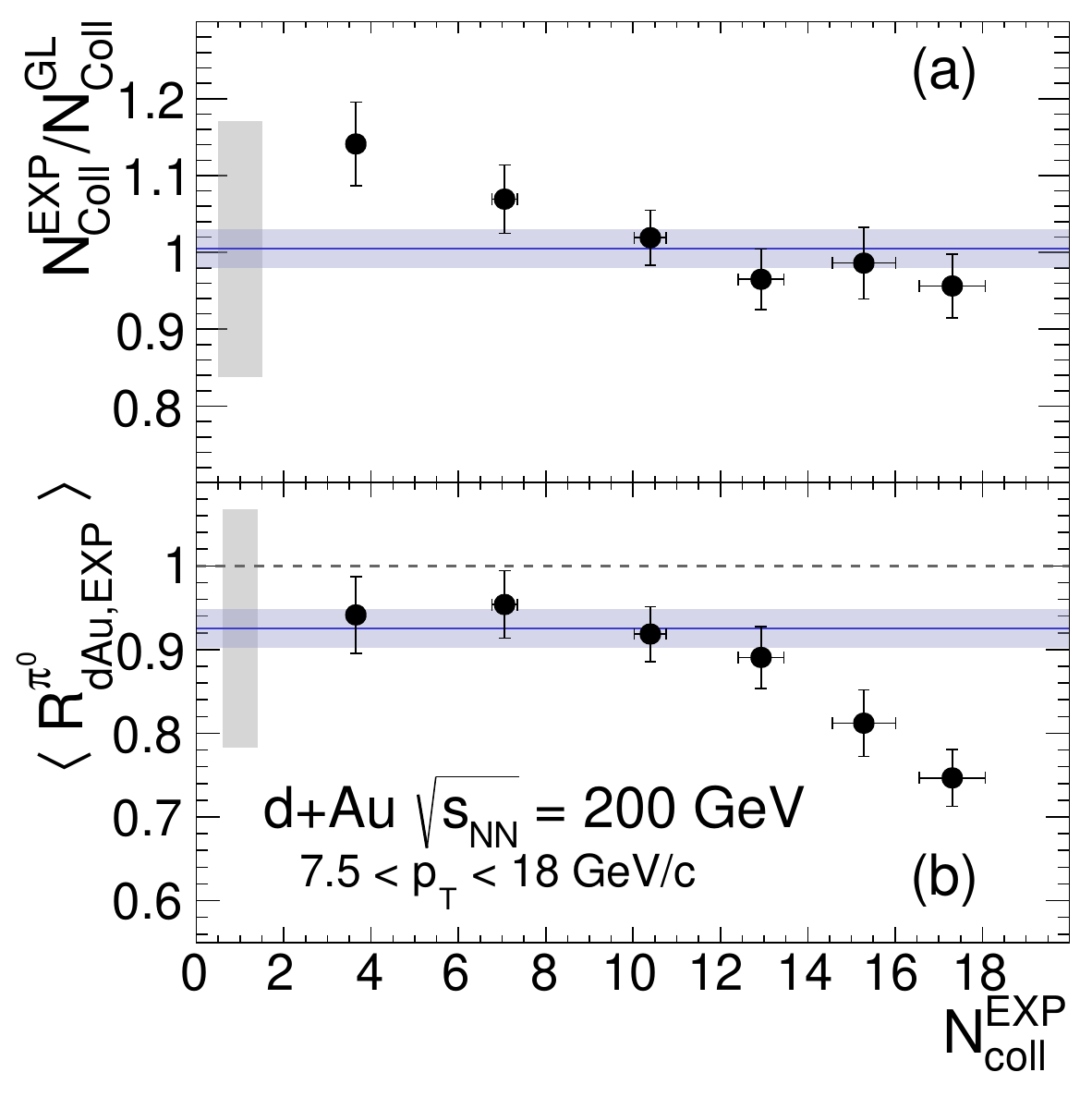}
\caption{The ratio (a) \Nexp/\Ngl and (b) the average \RdAuPiExp as a
function of \Nexp.  Horizontal and vertical bars are the statistical
uncertainties. The values for 0\%--100\% centrality \dau collisions are
represented by a solid [blue] line, with the statistical uncertainty
given as a band. The scale uncertainties that are common to all data
points are shown for the 0\%--100\% value.}
        \label{fig:rdau_cent}
\end{figure}

To further quantify the bias in mapping event activity to \Ngl, \Nexp and 
\Ngl are compared directly for the different event classes. 
Figures~\ref{fig:rdau_pt}(a) to (c) show \Nexp versus \pt for MB \dau 
events (0\%--100\%), and those with high (0\%--5\%) and low (60\%--88\%) 
event activity. The figure includes the average values of \Nexp 
determined from fits to the data (solid lines) compared to \Ngl (dashed 
lines)~\cite{PHENIX:2014nkk}. The systematic uncertainties on \Nexp, 
${\approx}16$\%, are dominated by uncertainties on the \pp data set and 
thus are a common scale uncertainty for all \dau event classes. The \Nexp 
and \Ngl agree well for 0\%--100\%, and are consistent for all event 
selections within uncertainties. However, the ratio of \Nexp/\Ngl shown 
in Fig.\ref{fig:rdau_cent}(a) has a clear trend with event activity. The 
ratio is larger than one for events with low activity. The \Nexp/\Ngl 
ratio decreases with increasing activity and becomes consistent with 
unity. The significance of this trend is evaluated by calculating the 
double ratio, $\Nexp($i$)/\Nexp(0\%$--100\%) to 
$\Ngl($i$)/\Ngl(0\%$--100\%) for the event selection $i$. For 0\%--5\% 
and 60\%--88\% the double ratio is $0.96\pm0.05 \pm 0.01$ and 
$1.16\pm0.07 \pm 0.06$, respectively. The systematic uncertainties mostly 
cancel so that only the uncertainties on the bias factor remain.  Because 
\Ngl and \Nexp agree reasonably well for 0\%--100\% and events with large 
event activity, it seems that \Ngl underestimates the number of hard 
scattering processes in events with low event activity. This may have led 
to the previously observed enhancement of \rxa for \piz in \pau, \dau and 
\heau collisions~\cite{PHENIX:2021dod}.

Next, possible nuclear modifications of \piz production in \dau collisions 
with high event activity are investigated.  For this the nuclear 
modification factor is calculated using \Nexp (as defined in 
Eq.~\ref{eq:Ncoll_exp}) instead of \Ngl,
\begin{equation}\label{eq:double_ratio}
  \RdAuPiExp  = \frac{\ypidau}{\Nexp \ \ypipp} 
             = \frac{\ydgpp/\ypipp}{\ydgdau/\ypidau},
\end{equation}
\noindent which is equivalent to the double ratio of \gdir/\piz ratios. 
Figures~\ref{fig:rdau_pt}(d) to (f) show \RdAuPiExp for the same event 
classes as panels (a) to (c). Over the observed \pt range there is no 
appreciable \pt dependence; the results of fits to the data are also 
indicated. Within uncertainties, \RdAuPiExp for 0\%--100\% is consistent 
with unity. The same is the case for \RdAuPiExp from lowest event-activity 
sample (60\%--88\%). In contrast, for the highest event activity-sample 
(0\%--5\%), a small but significant suppression of ${\approx}20$\% can be 
seen.

The evolution of the average \RdAuPiExp as a function of \Nexp is shown 
in Fig.~\ref{fig:rdau_cent}(b).  The points below 14 in \Nexp are consistent 
with the 0\%--100\% value, and within the scale uncertainty of 16.5\% 
consistent with unity or a few percent increase above unity, which would 
be expected from CNM effects~\cite{Arleo:2011gc}. However, for the 
collisions with the largest event activity \RdAuPiExp is significantly 
reduced. The reduction is quantified by a double ratio in which the 
systematic uncertainties cancel:
\begin{equation}
    \frac{\RdAuPiExp(0\%\mbox{--}5\%)}{\RdAuPiExp(0\%\mbox{--}100\%)} = 0.806\pm0.042,
\end{equation}
with a 4.5$\sigma$ deviation from unity. The same ratio for the events 
with the smallest event activity is $1.017\pm0.056$, consistent with 
unity. The observed $0.806\pm0.042$ suppression of the \piz yield 
corresponds to a \pt shift of $\delta\pt=0.213\pm0.055$ \gevc at 9 
\gevc. This shift is smaller than upper limits currently set by LHC 
experiments~\cite{ALICE:2017svf,ATLAS:2022iyq}. 
energy loss was limited to 0.4 \gevc outside of a cone of $R=0.4$ for 15 
\gevc jets from \pPb collisions~\cite{ALICE:2017svf}. These limits 
suggest that the LHC measurement would not be sensitive to the 
suppression observed here. Multiple factors increase the sensitivity of 
PHENIX, including (i) experimental techniques, such as  eliminating any model 
dependence, minimizing systematic uncertainties through double ratios, 
and choosing a larger system, (ii) the softer momentum spectrum at RHIC 
compared to at the LHC, and (iii) measuring leading particles rather than 
partial jet energies.


In summary, with the simultaneous measurement of \piz and \gdir at high 
\pt in \dau collisions at $\sqrt{s_{_{NN}}} = 200$ \gev, PHENIX has 
established that the previously observed enhancement of \piz $R_{dAu}$ 
in events with low activity is likely caused by an event-selection bias 
in estimating \Ngl within the GLM framework. The \Nexp based on direct 
photons, introduced in this paper, provides a more accurate 
approximation of the hard-scattering contribution. Using \Nexp 
eliminates the enhancement, while maintaining a 20\% suppression of high 
\pt \piz in events with high activity. The observed suppression is 
qualitatively consistent with the predictions of energy loss in small 
systems~\cite{Huss:2020whe,Ke:2022gkq}. If the suppression is indeed due 
to hot-matter effects, the yield of fragmentation photons within \gdir 
may also be suppressed, which in turn would lead to a slight 
underestimate of the suppression. To the contrary, any remaining bias in the 
event selection, for example due to the different Bjorken-$x$ range 
sampled by events with \gdir or \piz with the same \pt, could add to the 
suppression. Further studies of the system-size dependence with \pau, 
\dau, and \heau collisions may shed more light on the origin of the 
observed suppression. 





We thank the staff of the Collider-Accelerator and Physics
Departments at Brookhaven National Laboratory and the staff of
the other PHENIX participating institutions for their vital
contributions.  
We acknowledge support from the Office of Nuclear Physics in the
Office of Science of the Department of Energy,
the National Science Foundation,
Abilene Christian University Research Council,
Research Foundation of SUNY, and
Dean of the College of Arts and Sciences, Vanderbilt University
(U.S.A),
Ministry of Education, Culture, Sports, Science, and Technology
and the Japan Society for the Promotion of Science (Japan),
Natural Science Foundation of China (People's Republic of China),
Croatian Science Foundation and
Ministry of Science and Education (Croatia),
Ministry of Education, Youth and Sports (Czech Republic),
Centre National de la Recherche Scientifique, Commissariat
{\`a} l'{\'E}nergie Atomique, and Institut National de Physique
Nucl{\'e}aire et de Physique des Particules (France),
J. Bolyai Research Scholarship, EFOP, the New National Excellence
Program ({\'U}NKP), NKFIH, and OTKA (Hungary),
Department of Atomic Energy and Department of Science and Technology
(India),
Israel Science Foundation (Israel),
Basic Science Research and SRC(CENuM) Programs through NRF
funded by the Ministry of Education and the Ministry of
Science and ICT (Korea).
Ministry of Education and Science, Russian Academy of Sciences,
Federal Agency of Atomic Energy (Russia),
VR and Wallenberg Foundation (Sweden),
University of Zambia, the Government of the Republic of Zambia (Zambia),
the U.S. Civilian Research and Development Foundation for the
Independent States of the Former Soviet Union,
the Hungarian American Enterprise Scholarship Fund,
the US-Hungarian Fulbright Foundation,
and the US-Israel Binational Science Foundation.



%
 
\end{document}